\documentclass[11pt]{article}
\usepackage{amsfonts,amsbsy,amsmath,amssymb}

\newcommand{\BE}{\begin{equation}}
\newcommand{\EE}{\end{equation}}
\newcommand{\BA}{\begin{eqnarray}}
\newcommand{\EA}{\end{eqnarray}}

\newcommand{\msbar}{\overline{\rm MS}}
\newcommand{\msbarQ}{\overline{\rm MS}(\mu=Q)}

\begin{document}

\begin{titlepage}

\vspace*{22mm}
\begin{center}
              {\LARGE{\bf  Brodsky} {\it et al}'s {\bf defence does not work}}
\vspace{23mm}\\
{\large P. M. Stevenson}
\vspace{4mm}\\
{\it
T.W. Bonner Laboratory, Department of Physics and Astronomy,\\
Rice University, Houston, TX 77251, USA}

\vspace{30mm}

{\bf Abstract:}

\end{center}

\begin{quote}
In their defence of ``maximum conformality'' methods, Brodsky {\it et al} make the astonishing claim that 
any RG transformation $a'=a(1+V_1 a+\ldots)$ in QCD must have $V_1$ proportional to $b=(33-2 n_f)/6$.  
It is well known that this is not true.    I emphasize again the correctness and central importance of the  
Celmaster-Gonsalves relation for the prescription dependence of the $\Lambda$ parameter.

\end{quote}

\end{titlepage}

\setcounter{equation}{0}

    I recently made a direct attack \cite{MaxConDNW} upon the  BLM \cite{BLM} or PMC (``principle of maximum conformality'')  
methods (see \cite{gbrww} and references therein).  The key point arises already at next-to-leading order (NLO) for a physical 
quantity
\BE
\label{apgen}
{\cal R} = a(1+ r_1 a + \ldots).
\EE
The result given by  BLM/PMC for the $r_1$ coefficient  is just {\it different} when one starts, not from $\msbar$, 
but from another scheme related to it by 
\BE
\label{ap}
a' = a(1+ V_1 a + \ldots).  
\EE
Thus BLM/PMC does not resolve the scheme ambiguity \cite{MaxConDNW,CelStechyla}.  None of the variants or
generalizations of BLM cures this fundamental defect.  

    Brodsky {\it et al}'s defence \cite{BrodComm} goes off in many directions.  I disagree with almost every statement they make, but 
it would be of little wider interest to rebut each one.  The key point of their defence, though, is astonishing: They claim in 
their seventh paragraph that $V_1$ must be proportional to $b=(33-2 n_f)/6$ and that anything else would somehow spoil the 
``colour structure.'' 

    It is easily seen that this is untrue.   Ever since the classic work of Celmaster and Gonsalves (CG) \cite{CG} it has been clear 
that {\it scheme} dependence involves both {\it scale} and {\it prescription} dependence, so that a general {\it scheme} 
transformation has the form 
\BE
V_1 = - b \ln(\mu'/\mu) + v_1.
\EE
While the {\it scale}-change part is proportional to $b$, the {\it prescription}-change part, $v_1$, can be any finite number and in QCD
 could depend arbitrarily upon $n_f$  (though for most prescriptions in the literature $v_1$ is linear in $n_f$).  CG introduced a class of 
``momentum subtraction'' (MOM) prescriptions and calculated how they relate to minimal subtraction (MS) in NLO.  
Their results, in Eqs. (17, 18), are of the form of Eq.~(\ref{ap}) with $v_1$'s that are clearly not proportional to $33-2 n_f$.   
(CG also explicitly checked that their calculations satisfied the Ward identities.)  Do Brodsky {\it et al} really believe that all MOM 
prescriptions are illegitimate?  

    Further, if the ``$V_1 \propto b$'' contention were true, it would need to be explained why MS and $\msbar$ are correct, 
while a similar prescription \cite{CG} that defined the trace of the unit Dirac matrix in $d$ dimensions to be $d$ or $2^{d/2}$ (rather than $4$) 
would somehow be flat wrong.  

    The point can also be made as follows. One way to define a renormalization prescription (as noted by many authors) 
 is through the ``effective charge'' associated with a physical quantity.  For example, from the $e^+e^-$ cross-section ratio 
one could define a prescription whose couplant is related to the $\msbar$ one by a $V_1$ equal to the $r_1$ coefficient for 
${\cal R}^{e^+e^-}$ calculated in $\msbar$:  
\BE
r_1^{e^+e^-\,(\msbar)} = b \ln(\mu/Q) + \left( \frac{365}{24}- 11 \zeta_3 \right)  - n_f \left( \frac{11-8 \zeta_3}{12} \right),
\EE
which again, is {\it not} proportional to $b$.  Does the $e^+e^-$ effective-charge prescription ``violate the colour structure''?  
(Or does $\msbar$?)  Of course not.  

   The scale/prescription issue is one that still seems to confuse many researchers.  
The key to understanding it is the CG relation \cite{CG}:
\BE 
\label{CG0a}
\ln(\tilde{\Lambda}^{\prime}/\tilde{\Lambda}) = v_1/b,  
\EE 
which the relates the $\tilde{\Lambda}$ parameters of two prescriptions.  

   One must first define $\tilde{\Lambda}$ within a given renormalization prescription.  The $\beta$-function equation 
\BE
 \mu \frac{d a}{d \mu} \equiv \beta(a) = -b a^2(1+ c a + c_2 a^2 + \ldots) 
\EE
is integrable and (since we naturally want the range of integration to involve only $0$ to $a$) 
the constant of integration can be conveniently specified \cite{OPT} by\footnote{
For a full discussion see Sect. 6.3 of \cite{PMSbook}.  CG did not quite arrive at this way of defining a $\Lambda$ parameter, 
but their discussion in subsection II.F of \cite{CG} is completely consistent with it.   This definition of $\tilde{\Lambda}$ is not unique, 
but any satisfactory definition will be equivalent to it up to the addition of a finite constant to ${\cal C}(\delta)$; for example, the 
conventional definition of $\Lambda$ corresponds to subtracting 
$(c/b) \ln \mid\!2 c/b\!\mid$. 
}

\BE
\label{intbeta0}
\ln(\mu/{\tilde{\Lambda}}) =
\lim_{\delta \to 0} \left( \int_\delta^{a} 
\frac{d x}{ \beta(x)} + {\cal C(\delta)} \right), 
\EE
with
\BE
\label{Cdeltadef}
{\cal C}(\delta) = \int_\delta^{\infty} 
\frac{d x}{ b x^2(1 + c x)}.  
\EE

     The $\tilde{\Lambda}$ parameter depends upon the renormalization prescription, but in the simple and definite way given by Eq.~(\ref{CG0a}) above.  
That equation is {\it exact} and does not involve any $v_2, v_3, \ldots$ coefficients.  CG's proof, in my notation, is the following.  
In the primed prescription the counterpart to Eq.~(\ref{intbeta0}) will need primes on $\tilde{\Lambda}$, on $a$, and (beyond NLO) on $\beta(x)$.
(The $\mu$ remains the same since, by definition, a change of {\it prescription} is a change of {\it scheme} that keeps the value of $\mu$ unchanged.)  
One therefore has 
\BE
\label{intbetapr}
\ln(\mu/{\tilde{\Lambda}}') =
\lim_{\delta\to 0} \left( \int_{\delta}^{a'} 
\frac{d x}{ \beta'(x)} + {\cal C}(\delta) \right).  
\EE
When taking the difference of Eqs.~(\ref{intbeta0}) and (\ref{intbetapr}) the ${\cal C}(\delta)$ terms cancel leaving
\BE
\label{diff}
\ln(\mu/{\tilde{\Lambda}}) - \ln(\mu/{\tilde{\Lambda}}')  = 
\lim_{\delta \to 0} \left( \int_\delta^{a} 
\frac{d x}{ \beta(x)} -  \int_{\delta}^{a'} 
\frac{d x}{ \beta'(x)} \right), 
\EE
On the left-hand side the $\mu$ dependence cancels to leave $\ln(\tilde{\Lambda}^{\prime}/\tilde{\Lambda})$.  The right-hand side 
must therefore be {\it independent} of $\mu$, so we may {\it choose} to take $\mu \to \infty$, making both $a$ and $a'$ tend to zero, by 
asymptotic freedom.  One may then evaluate the difference of integrals for arbitrarily small $a$ values, which leads to the result in 
Eq.~(\ref{CG0a}) above.  

     CG's proof is sound, but it does leave a bit of a mystery as to how the right-hand side of Eq.~(\ref{diff}) reduces to 
$v_1/b$ at {\it any} $\mu$.  An alternative proof, due to Hugh Osborn, can be found in Sect.~6.5 of \cite{PMSbook}.  It splits the 
two integrations at $\epsilon$ and $\epsilon'= \epsilon(1+ v_1 \epsilon+\ldots)$, respectively.  Two terms cancel exactly, leaving only terms with $x<\epsilon$ 
or $x<\epsilon'$.  Thus it is clear that the non-zero result, $v_1/b$, arises solely from the $\frac{1}{x^2}$ singularity and the behaviour of 
$\beta(x)$ away from the infinitesimal neighbourhood of $x=0$ plays no role.\footnote{
Ref.~\cite{McKeon}, cited by \cite{BrodComm} in support of their claim, has tacitly assumed that $\Lambda$ is invariant, so the 
right-hand side of their equation (18) is missing a $\ln(\Lambda^*/\Lambda)$ term.  
}
\footnote{
   In ref. \cite{MaxConDNW} I said that ``${\cal C}(\delta)$ involves only $b$ and $c$ and so is RS invariant.''  The second part 
of that statement is problematic because in different prescriptions one may {\it choose} whether or not to use the same $\delta$.  
My proof of the CG relation in Appendix A of \cite{MaxConDNW} used a $\delta'=\delta(1+v_1 \delta + \ldots)$ and the $v_1/b$ 
result then arose from ${\cal C}(\delta) -{\cal C}(\delta')$.  Brodsky {it et al} seize upon this semantic conundrum, but it has no bearing 
on the correctness of the CG result.  
}

   The CG relation is the key to the observation \cite{OPT} that
\BE
\label{rho1Q}
{\boldsymbol \rho}_1(Q) \equiv b \ln(\mu/\tilde{\Lambda}) - r_1
\EE
is an invariant -- independent of both $\mu$ and the prescription.  That means that the NLO coefficient $r_1$ of a physical quantity ${\cal R}$ 
depends on scheme {\it only} through the ratio $\mu/\tilde{\Lambda}$  (which makes sense, since the $a$ obtained from Eq.~(\ref{intbeta0}) at 
NLO depends only on $\mu/\tilde{\Lambda}$.)  

 There is no meaningful answer to the questions 
``What is the right $\mu$?'' and ``What is the right prescription?''  The only RS variable that matters at NLO is the {\it ratio} 
of $\mu$ to the $\tilde{\Lambda}$ parameter, so the real question, for any given quantity ${\cal R}$,  is ``What is a good 
choice of $\mu/\tilde{\Lambda}$?''   Methods that ``work'' supply an answer to that question.  

   I stress again that here and in \cite{MaxConDNW} I am not concerned with defending my own ``Principle of Minimal Sensitivity'' 
method \cite{OPT}:  That is a separate argument.   I have explained that method in detail in Ref. \cite{PMSbook} and anyone who 
applies it carefully will find, I am sure, that the method defends itself. 

\newpage

\section*{Revised Addendum:  Numerical Example} 

\renewcommand{\theequation}{A.\arabic{equation}}
\setcounter{equation}{0}

This addendum has been modified to conform to the PMC procedure, as specified in the Appendix to version 2 of Ref.~\cite{BrodComm}.  
I had not appreciated that, even at NLO, PMC differs from the original BLM procedure.  The decomposition of $r_1$ into ``$A n_f  + B$'' 
is now more mysterious:  Apparently there are some $n_f$ terms that {\it are} $n_f$ terms, and other $n_f$ terms that are {\it not} 
$n_f$ terms, but are ``$n_f'$ terms'' and belong in the ``$B$'' term. (I will use $n_f^A, n_f^B$ rather than $n_f, n_f'$.)

The specific example here is for ${\cal R}_{e^+e^-}$ with $n_f=2$ at $Q=1$~GeV with $\tilde{\Lambda}_{\msbar}=0.2$~GeV,  
corresponding to Example 3 ($Q/\tilde{\Lambda}_{\msbar}=5$) in Chap.~10 of Ref.~\cite{PMSbook}.   Everything up to 
Eq. (\ref{PMCresult}) below now agrees with their numbers.  My main point -- that $C_1^* \equiv \frac{33}{2}A+B$, and hence 
the PMC result, is RP dependent -- is unaffected.

     The NLO $\beta$ function is 
\BE
\beta^{{\scriptscriptstyle{\rm NLO}}}(a) = - b a^2(1+ c a),
\EE
with 
\BE
b= 29/6, \quad\quad c= 115/58.
\EE
The $r_1$ coefficient of the $\msbarQ$ scheme is decomposed\footnote{
I suspect that this decomposition depends on the wavefunction renormalization, but let that pass.} 
as $r_1=A n_f^A +B$ with 
\BA
A & = & -\frac{55}{72}+ \frac{2}{3} \zeta_3 = 0.0374824 , \nonumber \\
B & = & \frac{365}{24}-11 \zeta_3 - \frac{11}{72} n_f^B = 1.98571 -0.152778 \, n_f^B,
\EA
so that for $n_f=2$ (that is, $n_f^A=n_f^B=2$) one has 
\BE
r_1^{\msbarQ}=2A+ \left.  B \right|_{n_f=2}=1.75512.
\EE

    The NLO result in the $\msbarQ$ scheme uses an ``$a$'' that is the solution to the integrated $\beta$-function equation 
at NLO for $\ln(\mu/\tilde{\Lambda})$ equal to $\ln 5$; i.e., it is the root of 
\BE
\label{intbeta1}
\frac{1}{a} \left( 1+ c a \ln \left| \frac{c a}{1+ c a} \right| \right) - b \ln 5 =0,
\EE
which is 
\BE
a_{\msbar}(Q) = 0.0862557.
\EE
The NLO result in the $\msbarQ$ scheme is then
\BE
{\cal R}_{\msbarQ} = a_{\msbar}(Q) \left( 1+ r_1^{\msbarQ} a_{\msbar}(Q) \right) = 0.0993138.
\EE

    To obtain the PMC result I follow their instructions.  The PMC $r_1$ coefficient (which they now call $r_{1,0}$) is
\BE
r_{1,0} \equiv C_1^* \equiv \frac{33}{2}A + B = 2.29861,
\EE
and the PMC scale is 
\BE
\mu_r^*=Q \exp{3 A} = 1.11901 \, {\rm GeV}.
\EE
To find the PMC couplant, which is $a_*=a_{\msbar}(\mu\!=\!\mu_r^*)$, one should not use the 
LO formula, Eq. (3.4) of \cite{gbrww} (as it seems they now agree), but should solve the NLO integrated $\beta$-function equation, 
Eq. (\ref{intbeta1}) above with $b \ln5$ replaced by $b \ln(\mu_r^*/(0.2 \, {\rm GeV}))$.  That gives
\BE 
a_* = 0.0817833.
\EE
The PMC result for ${\cal R}$ at NLO is then
\BE
\label{PMCresult}
{\cal R}_{PMC}= a_*(1+C_1^* a_*) = 0.0971576.
\EE
(This is the $\msbar$ result but with $\mu= 1.11901 \, {\rm GeV}$ rather than $\mu=Q=1 \, {\rm GeV}$; a fact that 
seems to surprise them.)

   But suppose we had calculated $r_1$ in a different RP.  That is equivalent to taking 
the $\msbar$ result and making a change of RP, $a'= a (1 + v_1 a + ...)$.  (Let us choose to 
keep the scale choice as $\mu=Q$ in both prescriptions.)  The NLO coefficient in the new scheme is
\BE
{r_1}'  =  r_1 - v_1, 
\EE
and by the exact CG relation we have 
 \BE
{\tilde{\Lambda}}' =  \tilde{\Lambda}_{\msbar} e^{v_1/b}.  
\EE
(Both my examples below have $v_1=2$ so that ${\tilde{\Lambda}}'$ will be $0.302509$ GeV.)

      I contend that $v_1$ can depend on $n_f$ in any way whatsoever.    (Different values of $n_f$ give distinct 
theories describing distinct hypothetical universes and in each theory we may choose any value of $v_1$.)  However, 
I shall confine myself to considering $v_1$'s  linear in $n_f$.  That is,
\BE
v_1 =  v_{10}\, n_f^A + v_{11}.
\EE
In keeping with the new theology I have specified that $n_f$ here is an $n_f^A$ term (any $n_f^B$ term would effectively 
be absorbed in the $v_{11}$ term).  

\newpage

\noindent   {\bf Example (i)}:  $v_{10}=0$, $v_{11}=2$.  

    For brevity I skip the details and just state that this leads to 
\BE
{C_1^*}'= 0.298611,  \quad\quad {a_*}'= 0.101345  ,
\EE 
and hence
\BE
{{\cal R}_{PMC}}'= 0.104412.   .
\EE
The $C_1^*$ coefficient has changed, and so has the PMC result.

\vspace*{2mm}

\noindent  {\bf Example (ii)}:  $v_{10}=1$, $v_{11}=0$.  

       In the new RP we have
\BE
A'=A-1=-0.962518 , \quad B'=B=1.68015,
\EE
so that
\BE
{C_1^*}'= \frac{33}{2}A'+B'=-14.2014,
\EE
\BE
{\mu_r^*}'=Q \exp{3 A'}=\mu_r^*\, e^{-3} = 0.0557124 \, {\rm GeV}. 
\EE
Then $a_*'$ is the root of 
\BE
\frac{1}{a} \left( 1+ c a \ln \left| \frac{c a}{1+ c a} \right| \right) - b \ln  \left(\frac{{\mu_r^*}'}{{\tilde{\Lambda}}'} \right) =0.  
\EE
which gives 
\BE
{a_*}'=-0.15311
\EE
(there is no positive root).  Hence,  
\BE
{{\cal R}_{PMC}}' =  {a_*}' (1+ {{C_1}^*}' {a_*}') = -0.486028.  
\EE
Note that we are the wrong side of a Landau pole and the couplant  is negative.  

        In their version of these examples their $r_1^{\rm{\bf I}}$ coefficient (my $r_1'$) is written in Eq.~(27) as 
\BE
{\mbox{\rm``}} \,  r_1^{\rm{\bf I}}=r_1 - r_0 v_1=r_{1,0}+r_{1,1} b - r_0 v_1, {\mbox{\rm''}}
\EE
where ``$r_{1,0}$'' is a new name for $C_1^*$, ``$r_{1,1}$'' is not defined but must be $-3A$, and ``$r_0$'' is $1$.  That is, 
it is
\BE
r_1'= C_1^* - 3 A b - v_1
\EE
They then assert that that the $-v_1$ term ``should be treated as non-conformal term,'' meaning, apparently, that it must be treated 
as if it were proportional to $\frac{33}{2}-n_f^A$, even though it isn't.   Their ``proof''  that $C_1^*$ remains unchanged thus depends upon 
assuming the very result it purports to prove!  

      Example (ii), with $v_1 \propto n_f^A$, can arise very naturally in dimensional regularization.  As pointed out first by CG (Ref. \cite{CG}, 
last paragraph of section III), one might well modify the usual $\msbar$ rules by setting the trace of the unit Dirac matrix in $4-\epsilon$ dimensions 
to be $4(1-\sigma \epsilon+\ldots)$, where $\sigma$ is some numerical coefficient.  The $n_f$ part of the $b/\epsilon$ divergent term in the bare $r_1$ would 
then produce an extra finite contribution, $\frac{1}{3} \sigma n_f^A$.  (Thus, $\sigma$ is trivially related to my $v_{10}$.)  Note that this extra term 
is unquestionably an ``$n_f^A$'' term, as it arises from a divergent diagram contributing to the coupling renormalization.   I have mentioned this 
point repeatedly; it is telling that Brodsky {\it et al} never respond to it.

\end{document}